\documentclass[%
 aip,
 amsmath,amssymb,
 reprint,%
]{revtex4-1}

\usepackage{graphicx}
\usepackage{dcolumn}
\usepackage{bm}

\usepackage[utf8]{inputenc}
\usepackage[T1]{fontenc}
\usepackage{mathptmx}

\usepackage[version=4]{mhchem}
\usepackage{physics}
\usepackage[table,usenames,dvipsnames]{xcolor}
\usepackage{tabularx}
\usepackage[para]{threeparttable}
\usepackage{enumerate}
\usepackage{nicefrac}
\usepackage{natmove}
\usepackage{rotating} 

\usepackage{ulem}

\newcommand{\ib}{\bar{i}}
\newcommand{\jb}{\bar{j}}
\newcommand{\ab}{\bar{a}}
\newcommand{\bb}{\bar{b}}
\newcommand{\re}{$r_\mathrm{e}$}
\newcommand{\we}{$\omega_\mathrm{e}$}
\newcommand{\de}{$D_\mathrm{e}$}
\newcommand{\Ta}{$T_\mathrm{e}$}
\newcommand{\Tv}{$T_\mathrm{v}$}

\usepackage{soul}

\begin{document}

\preprint{AIP/123-QED}

\title[]{A configuration interaction correction on top of pair coupled cluster doubles}%

\author{Artur Nowak}
\affiliation{
Institute of Physics, Faculty of Physics, Astronomy and Informatics,
Nicolaus Copernicus University in Toru\'n, Grudziadzka 5, 87-100 Toru\'n, Poland
}%

\author{Katharina Boguslawski}
\email{k.boguslawski@fizyka.umk.pl}
\affiliation{
Institute of Physics, Faculty of Physics, Astronomy and Informatics,
Nicolaus Copernicus University in Toru\'n, Grudziadzka 5, 87-100 Toru\'n, Poland
}%

\date{\today}

\begin{abstract}
Numerous numerical studies have shown that geminal-based methods are a promising direction to model strongly correlated systems with low computational costs.
Several strategies have been introduced to capture the missing dynamical correlation effects, which typically exploit \textit{a posteriori} corrections to account for correlation effects associated with broken-pair states or inter-geminal correlations.
In this article, we scrutinize the accuracy of the pair coupled cluster doubles (pCCD) method extended by configuration interaction (CI) theory.
Specifically, we benchmark various CI models, including, at most double excitations against selected CC corrections as well as conventional single-reference CC methods.
A simple Davidson correction is also tested.
The accuracy of the proposed pCCD-CI approaches is assessed for challenging small model systems such as the \ce{N2} and \ce{F2} dimers and various di- and triatomic actinide-containing compounds.
In general, the proposed CI methods considerably improve spectroscopic constants compared to the conventional CCSD approach, provided a Davidson correction is included in the theoretical model. At the same time, their accuracy lies between the linearized frozen pCCD and frozen pCCD variants.
\end{abstract}

\maketitle

\section{Introduction}

The reliable and efficient description of electron correlation effects is an open problem in quantum chemistry.
Although the correlation energy is quantitatively small compared to the total electronic energy, the proper description of the correlated motion of the electrons is crucial to understand physical and chemical phenomena like bond-breaking processes.
Typically, electron correlation effects are divided into a dynamical and a non-dynamical/static part.
To capture both contributions simultaneously, we require advanced methods that exploit a multireference description.
Examples thereof are multireference coupled cluster approaches \cite{bogus_mrcc,monika_mrcc, adamowicz-mrcc}, the density matrix renormalization group (DMRG) algorithm \cite{white,white2,white-qc,ors_springer,marti2010b,chanreview,ors_ijqc, hachmann_h50, dmrg_kurashige, yanai-review, baiardi-reiher, gunst-dmrg, ma-dmrg}, or quantum Monte Carlo methods \cite{qmc,qmc-book-chapter}.
Despite their high accuracy, these methods are computationally rather expensive.

An alternative strategy to account for strong electron correlation effects is to exploit geminal-based methods.~\cite{geminal-review}
The key idea of geminal-based electronic structure approaches is to use non-interacting electron pairs as the fundamental building blocks of the electronic wave function.
Various numerical studies using geminal-based methods, such as the antisymmetric product of interacting geminals \cite{coleman_1965,bratoz1965,silver_1969,silver_1970,apig-1,apig-2}, the antisymmetric product of strongly orthogonal geminals (APSG) \cite{hurley_1953,parks_1958,kapuy1966,kutzelnigg_1964,surjan-bond-1984,surjan-bond-1985,surjan-bond-1994,surjan-bond-1995,surjan_1999,surjan-bond-2000,surjan2012}, singlet-type strongly orthogonal geminals,~\cite{rassolov2002,rassolov2004,rassolov-sspg} geminals constructed from Richardson–Gaudin states,~\cite{johnson_2013, johnson2017strategies, johnson2020richardson, fecteau2021richardson, johnson-near-exact-2022} and the antisymmetric product of 1-reference orbital geminals \cite{limacher_2013,oo-ap1rog}, also known as the pair-coupled cluster doubles (pCCD) ansatz \cite{tamar-pcc}, yield promising results. 
Specifically, the pCCD approach is a product of geminal creation operators $\phi_{i}^\dagger$ acting on some vacuum state
\begin{equation}\label{eq:ap1rog}
         |\textrm{pCCD} \rangle = \prod_{i=1}^{P}\phi_{i}^\dagger | 0\rangle, 
\end{equation}
with $P$ being the number of electron pairs and
\begin{equation}\label{eq:phi}
\phi_{i}^\dagger =a^\dagger_i a^\dagger_{\bar{i}} + \sum_{a}^{\rm virt}c_{i}^{a}a^\dagger_a a^\dagger_{\bar{a}},
\end{equation}
where the sum runs over all virtual orbitals $a$ and $\{c_i^a\}$ are the geminal expansion coefficients.
In {eq.~\eqref{eq:phi}}, $i$ ($\bar{i}$) indicates spin-up (spin-down) electrons.
Using an exponential ansatz \cite{limacher_2013}, the pCCD wave function can be rewritten as
\begin{align}\label{eq:pccd}
|\textrm{pCCD} \rangle  	&= \exp (\sum_{i}^{\rm occ} \sum_{a}^{\rm virt} c_i^a a^\dagger_a a^\dagger_{\bar{a}} a_{\bar{i}} a_i) | \Phi _0\rangle  \nonumber \\
													&= \exp(\hat{T}_\textrm{p})\ket{\Phi_0},
\end{align}
where $|\Phi_0 \rangle$ is some independent-particle wave function, for instance, the Hartree--Fock (HF) determinant, and $\hat{T}_\textrm{p}$ is the electron-pair cluster operator. 
One benefit in using the exponential form is the proper (linear) scaling of the method with the number of electrons (size-extensivity).
To recover size-consistency it is necessary to apply an orbital-optimization protocol \cite{oo-ap1rog,tamar-pcc,ps2-ap1rog,ap1rog-jctc} which also leads to localized orbitals, which are symmetry-broken \cite{oo-ap1rog,tamar-pcc,ps2-ap1rog,ap1rog-jctc}.
Investigations on the one-dimensional Hubbard model \cite{oo-ap1rog,boguslawski2016} or molecules with stretched bonds \cite{pawel_jpca_2014,tamar-pcc,frozen-pccd,boguslawski2015,ijqc-eratum,pawel_pccp2015,garza2015actinide,kasia-lcc,AP1roG-PTX,filip-jctc-2019,state-specific-oopccd}, even those containing lanthanide \cite{pawel-yb2} or actinide atoms \cite{pawel_pccp2015,garza2015actinide,nowak-pccp-2019} demonstrate that geminal-based approaches can capture an important part of static/nondynamic electron correlation effects.
Since pCCD requires a reasonable amount of computational resources (with a scaling of $N^4$ or $N^3$ if proper intermediates are defined), heavy-element-containing compounds are a promising scope of applications.~\cite{ola-book-chapter-2019} 
Nevertheless, by restricting the wave function ansatz to electron-pair states, a large fraction of the correlation energy is not accounted for.
This missing dynamical electron correlation energy that cannot be described using electron-pair states, is commonly included \textit{a posteriori} by means of, for instance, perturbation theory~\cite{rosta2002, jorgensen-ch+,piotrus_pt2,AP1roG-PTX}, a coupled cluster ansatz\cite{frozen-pccd,zoboki2013,kasia-lcc,leszczyk-jctc}, or density functional approximations~\cite{garza-pccp,garza2015}.
In this work, we develop a different \textit{a posteriori} correction for a pCCD reference function using configuration interaction (CI) theory.
Specifically, we will focus on a CI correction, which is restricted to hole-particle excitations with respect to a reference determinant.
Similar CI models have been already combined with, for instance, an APSG~\cite{Kallay1},  generalized valence bond~\cite{exgem-ijqc-1987, gvb-pp-ci-jcp-1988, gvb-pp-ci-ijqc-1999, exgem-review-1999}, or an antisymmetrized geminal power~\cite{agp-ci} reference function.

This work is organized as follows. In section~\ref{sec:ci}, we briefly summarize the theory of the CI corrections on top of pCCD, which are introduced using a spin-free formalism.
The computational details are summarized in section~\ref{sec:details}, while the numerical results are presented in sections~\ref{sec:results}--\ref{sec:exstates}.
Finally, we conclude in section~\ref{sec:conclusions}.

\section{\label{sec:ci}CI Corrections with a pCCD Reference Function}
Conventional (truncated) CI methods define the electronic wave function as a linear combination of all possible ``excited'' configurations with respect to some reference configuration $\ket{\Phi_0}$,
\begin{align}\label{eq:ci}
\ket{\textrm{CI}}  =  \sum_{\mu} c_{\mu} \hat{\tau}_\mu \ket{\Phi_0}
                   =  \sum_{\mu} c_{\mu} \ket{{\Phi_\mu}},
\end{align}
where {in this compact representation} $c_{\mu}$ are the CI coefficients and $\hat{\tau}_\mu$ is an excitation operator including occupied--virtual (or hole--particle) excitations up to a predefined level.
To arrive at a CI-type correction, we replace the reference determinant with a pCCD reference function eq.~\eqref{eq:pccd},
\begin{align}\label{eq:pccd-ci}
|\textrm{pCCD-CI} \rangle  =  \sum_{\mu}c_{\mu}\hat{\tau}_{\mu} e^{\hat{T}_{\rm p}} |{\rm \Phi_0} \rangle.
\end{align}
Substituting this ansatz for the electronic wave function $\ket{\Psi}$ in the Schr\"odinger equation $\hat H \ket{\Psi} = E \ket{\Psi}$, we obtain
\begin{align}\label{eq:h-pccd-ci-1}
 \sum_{\mu} \hat{H} c_{\mu}\hat{\tau}_{\mu} e^{\hat{T}_{\rm p}}  |{\rm \Phi_0} \rangle 
    &= E \sum_{\mu}c_{\mu}\hat{\tau}_{\mu} e^{\hat{T}_{\rm p}}  |{\rm \Phi_0} \rangle \nonumber \\
 \sum_{\mu} \hat{H} e^{\hat{T}_{\rm p}}  c_{\mu}\hat{\tau}_{\mu} |{\rm \Phi_0} \rangle 
    &= E \sum_{\mu} e^{\hat{T}_{\rm p}}  c_{\mu}\hat{\tau}_{\mu} |{\rm \Phi_0} \rangle
\end{align}
where $\hat H$ is the electronic Hamiltonian comprising one- and two-electron integrals and we utilized that $\hat \tau_\mu$ and $\hat T_{\rm p}$ commute as they contain only hole--particle excitation operators, $[\hat \tau_\mu, \hat T_{\rm p} ] = 0$.
Unfortunately, the eigenvalue problem of eq.~\eqref{eq:h-pccd-ci-1} cannot be solved in its symmetric representation as the corresponding series expansion, that enters the working equation, does not truncate.
In order to solve for the eigenvalues and eigenvectors efficiently, we can project out the components of interests, which will lead to a diagonalization problem of a non-Hermitian matrix.
There are several choices for the projection space, which is defined in terms of the de-excitation operator $\hat \tau ^\dagger_\mu$.
We can either project by $\bra{\Phi_0} \hat \tau ^\dagger _\mu$ or $\bra{\Phi_0} \hat \tau ^\dagger _\mu e^{-\hat{T_{\rm p}}}$.
We should note that if we restrict the CI operator to comprise at most single and double excitations, both projection schemes will result in the same solutions.\cite{Kallay1}
In the following, we will focus on the latter approach.
Multiplying eq.~\eqref{eq:h-pccd-ci-1} from left with $e^{-\hat{T_{\rm p}}}$ and using the property that $\hat{T_{\rm p}}$ and $\hat{\tau}_\mu$ commute, results in
\begin{align}\label{eq:h-pccd-ci-2}
e^{-\hat{T_{\rm p}}}\hat{H} e^{\hat{T_{\rm p}}} \sum_{\mu}c_{\mu}\hat{\tau}_{\mu}  |{\rm \Phi_0} \rangle &= E \sum_{\mu}c_{\mu}\hat{\tau}_{\mu} e^{-\hat{T_{\rm p}}} e^{\hat{T_{\rm p}}} |{\rm \Phi_0} \rangle \nonumber \\
\mathcal{{H}}^{\rm (p)} \sum_{\mu}c_{\mu}\hat{\tau}_{\mu}  |{\rm \Phi_0} \rangle &= E \sum_{\mu}c_{\mu}\hat{\tau}_{\mu}  |{\rm \Phi_0} \rangle.
\end{align}
Thus, the complex CI problem has now the form of a conventional single-reference (truncated) CI model, where the Hamiltonian is substituted by the similarity-transformed Hamiltonian of pCCD $\mathcal{{H}}^{\rm (p)}$.
The diagonalization problem becomes computationally feasible and can be solved in efficient time as the commutator expansion of $e^{-\hat{T_{\rm p}}}\hat{H} e^{\hat{T_{\rm p}}}$ naturally truncates.
The corresponding working equations bear similarities to the EOM formalism~\cite{eomcc_1968,eomcc_1989,kowalski-eom-review-2011,krylov-review} as EOM-CC diagonalizes the similarity-transformed Hamiltonian of a given CC model.
The main difference between conventional EOM-CC and the proposed pCCD-CI model is that the similarity-transformed Hamiltonian to be diagonalized has a particular simple form (as it only accounts for electron-pair excitation), while the CI ansatz includes any choice for the excitation operator.
In standard EOM-CC theory, both the CC cluster operator and the CI ansatz are restricted to the same order of excitations.

Furthermore, we will consider only singlet excitations.
The corresponding projection manifold will result in a spin-free CI problem, where all eigenstates of $\mathcal{{H}}^{\rm (p)}$ will be singlet states.
Moreover, from all working equations, we have subtracted the energy of the reference function, that is, the pCCD total energy (similar to the conventional CI equations, where the energy of the reference determinant is subtracted).
Thus, the eigenenergies correspond to excitation energies with respect to the pCCD reference function.
In this spin-free formulation, the excitation operator that contains single excitations can be expressed as follows
\begin{align}\label{eq:sf-pccd-cis}
\hat{\tau} =\hat{\tau}_{0} + \hat{\tau}_{\rm S} 
    &= c_0 + \sum_{ia}c_i^a(a_a^{\dagger} a_i + a_{\ab}^{\dagger}a_{\ib} ) \nonumber \\
    &= c_0 + \sum_{ia}c_i^a E_i^a, 
\end{align}
where we introduced the singlet one-electron excitation operator $E_i^a =  a_a^\dagger a_i + a_{\ab}^\dagger a_{\ib}$.
To account for double excitations, the above operator is modified as follows
\begin{align}\label{eq:sf-pccd-cid}
\hat{\tau} &= \hat{\tau}_{0} + \hat{\tau}_{\rm D}  \nonumber \\
    &= c_0 + \frac{1}{2} \sum_{ijab}c_{ij}^{ab}(a_a^{\dagger}a_b^{\dagger} a_ja_i + a_a^{\dagger}a_{\bb}^{\dagger} a_{\jb} a_i + a_{\ab}^{\dagger} a_b^{\dagger} a_j a_{\ib}^{\dagger} + a_{\ab}^{\dagger}a_{\bb}^{\dagger} a_{\jb} a_{\ib} ) \nonumber \\
    &= c_0 + \frac{1}{2} \sum_{ijab} c_{ij}^{ab} E_i^a E_j^b. 
\end{align}
To arrive at a CISD correction, the excitation operators for singles and doubles are combined, that is, $\hat{\tau} = \hat{\tau}_{0} + \hat{\tau}_{\rm S} + \hat{\tau}_{\rm D}$.
We should note that we tested two different $\hat{\tau}_{\rm D}$ excitation operators.
In the pCCD-CID and pCCD-CISD model, all electron pair excitations are excluded from the CI ansatz as they are already accounted for in the pCCD reference calculations.
If we are interested in targeting several roots, that is, the electronic ground state and some lowest-lying electronically excited states, this approximation will neglect any double excitations of electron-pair character and, hence, bi-excited states might miss important contributions to the wave function expansion.
Thus, in the pCCD-CID$^\prime$ and pCCD-CISD$^\prime$ models, the electron-pair sector is added to the general excitation operator $\hat{\tau}_{\rm D}$.
However, including electron-pair excitations in the CI ansatz might lead to double-counting problems of electron correlation effects associated with electron-pair states.
In the following, we assess both CI ans\"atze (with and without electron-pair excitations) in describing ground and electronically excited states.

In order to solve eq.~\eqref{eq:h-pccd-ci-2}, that is, to find the eigenvalues and eigenvectors of $\mathcal{{H}}^{\rm (p)}$, its matrix representation has to be diagonalized.
In practical calculations, iterative techniques are used to compute only a few of the lowest eigenvalues and corresponding eigenvectors, for instance non-Hermitian extensions of the Davidson algorithm (the corresponding working equations are collected in the SI).
We should note that the projection manifold has a special form \cite{helgaker_book} (similar to spin-free CC theory) to allow for an algebraic spin summation, which results in the spin-free pCCD-CI working equations,
\begin{align}
    \bra{\overline{\Phi_I^A}}          &= \frac{1}{2} \bra{\Phi_0} E_{a}^i \label{eq:projection-s}\\
    \bra{\overline{\Phi_{IJ}^{AB}}}    &= \frac{1}{3} \bra{\Phi_0} E^j_{b}E^i_{a}  + \frac{1}{6} \bra{\Phi_0} E^j_{a} E^i_{b}.\label{eq:projection-d}
\end{align}

Finally, we include a Davidson correction on top of the proposed pCCD-CI models to minimize the size-consistency error intrinsic to (truncated) CI methods.~\cite{scc-overview,szalay2012}
Specifically, we combine a renormalized Davidson correction introduced in Ref.~\citenum{luken1978} exploiting the pCCD-CI correlation energy,
\begin{equation}
    E_{\rm{RDC}} = \Big(\frac{1-c_0^2}{c_0^2}\Big)\Big({E_{\rm pCCD-CI} - {E_{\rm RF}}}\Big),
\end{equation}
where  ${E_{\rm pCCD-CI}}$ indicates the total energy of the pCCD-CI method, ${E_{\rm RF}}$ is the energy of the reference method (here pCCD), and $c_0$  is the contribution of the reference determinant ($\ket{\Phi_0}$) of the reference method.

\section{\label{sec:details}Computational details}
\subsection{Basis sets, relativistic effects, and frozen core}
For the light diatomic molecules $\ce{H_2}$ and $\ce{N_2}$, we used the triple-$\zeta$ correlation consistent basis sets of Dunning~\cite{basis_dunning} (cc-pVTZ), while for $\ce{F_2}$ the augmented version (aug-cc-pVTZ) was employed.
For all actinide compounds, we used the double-$\zeta$ correlation consistent basis sets of Peterson~\cite{cc-pvdz-dk3-actinides} (cc-pVDZ-DK3) for all heavy elements, optimized specifically for the DKH3 Hamiltonian,~\cite{reiher_2004a,reiher_2004b,reiher_book,tecmer2016} and Dunning's aug-cc-pVDZ basis set for all the remaining light elements. 
Scalar relativistic effects were accounted for by the DKH3 Hamiltonian.
All calculations for systems containing heavy-elements were performed with a frozen core to ensure a compromise between computational efficiency, the reliability of the results for heavy-element-containing systems, and the quality of the atomic basis sets~\cite{real09,tecmer2014,pawel_pccp2015,ptThS}.
Specifically, the atomic 1s orbitals of the O and N atom, the 1s, 2s, and 2p orbitals of the S atom, and up to the 5d orbitals of the Th, Pa, and U center were frozen.
All post-pCCD methods as well as all CCD and CCSD calculations were performed using a developer version of the \textsc{PyBEST} v1.2.0 software package~\cite{pybest-paper,pybestv1.2.0,pybest_web}.
All pCCD-based calculations exploited the natural orbitals obtained from orbital-optimized pCCD, while all conventional CC calculations featured canonical Hartree--Fock orbitals.
The CCSD(T) reference calculations were performed in the \textsc{Molpro2020}~\cite{molpro} software suite.   


\subsection{Evaluations of spectroscopic constants}
The potential energy curves were obtained from a polynomial fit of 8-th order using the fitting scripts available in the \textsc{PyBEST}~\cite{pybestv1.2.0} software package.
From these fitted potential energy curves, we derived the corresponding spectroscopic constants such as equilibrium bond lengths ($r_{\rm e}$) and harmonic vibrational frequencies ($\omega_{\rm e}$) for all actinide compounds and light diatomic molecules.
Additionally, the potential energy well depths ($D_e$) were calculated from fitting a generalized Morse function.
To determine the harmonic vibrational frequencies ($\omega_{\rm e}$), we performed numerical calculations using the five-point finite difference stencil~\cite{Abramowitz} with the following averaged masses: uranium: 238.0508, thorium: 232.0381, protactinium: 231.0359, oxygen: 15.9949, sulfur: 31.9721, and nitrogen: 14.0031.~\cite{spectro-data-2005}



\begin{figure*}[t]
\hspace{-0.5cm}
	\includegraphics[width=2.0\columnwidth]{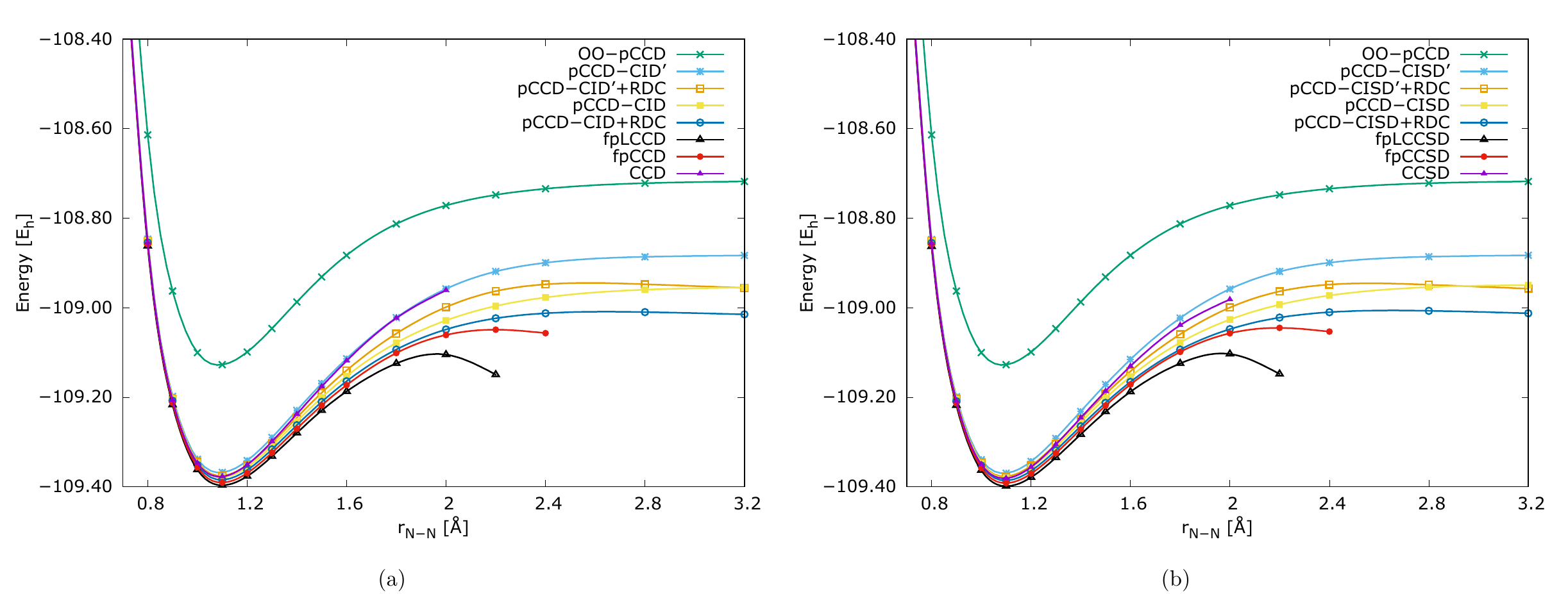}
	\tiny 
	\caption{Potential energy surfaces of the ground states of \ce{$N_2$} determined for various post-pCCD with (a) doubles excitation and (b) singles and doubles excitations. The superscript $^\prime$ indicates that the CI operator also includes the electron pair sector. RDC: renormalized Davidson correction.
	}
  	\label{fig:fig1}
\end{figure*}


\section{\label{sec:results}Assessing the accuracy of the CI corrections for light diatomic molecules}

Our first test set to assess the performance of the CI corrections contains the \ce{N2} and \ce{F2} diatomic molecules.
In these systems, the different flavors of electron correlation effects are difficult to describe with various pCCD corrections.
For instance, a remarkable challenging problem for computational chemistry is to accurately model the electronic structure of the \ce{N_2} molecule~\cite{chan_n2,entanglement_letter,frozen-pccd,boguslawski2017, nowak-jcp,leszczyk-jctc} along its dissociation pathway.
This problem is related to dynamic changes in the contributions of electron correlation effects for an increasing interatomic distance.
In general, around the equilibrium distances, dynamical electron correlation effects are prevalent, while the contribution of nondynamic/static electron correlation significantly increases when the N--N bond is stretched.
Furthermore, all CC corrections on top of pCCD restricted to at most double excitations fail and do not yield a physically meaningful dissociation limit.
Unlike in the nitrogen dimer, dynamical correlation effects are dominant in the \ce{F_2} molecule.
Conventional methods require even triple excitations to model the
dissociation pathway of \ce{F_2} reliably\cite{kowalski-f2_2001,boguslawski2017,nowak-jcp, leszczyk-jctc}.
Thus, the \ce{F_2} dimer represents a good test system to assess if novel methods can accurately describe a bond-breaking process without increasing the computational cost related to the inclusion of triple (or higher) excitations.
Since pCCD-based methods do not provide sufficiently accurate results in these systems, they represent a good testing ground to assess the accuracy and robustness of the CI corrections.

\begin{table*}[t]	
\centering
\caption{Spectroscopic parameters for the ground state of the \ce{N_2} molecule obtained from various post-pCCD methods and conventional CC models. \re\ is the equilibrium bond length, \we\ the harmonic vibrational frequency, and \de\ the dissociation energy, respectively.
The differences with respect to experimental result $\Delta_{\rm{exp}}$ or MRCI reference data $\Delta_{\rm{MRCI}}$ are given in parentheses. The total electronic energies are summarized in the SI.
RDC: renormalized Davidson correction.
} 

\label{tbl:n2}

{\footnotesize

\begin{tabular}{l|ccc|ccc|ccc}

	Method & \re\ [\AA]  & $\Delta_{\rm{MRCI}}$  & $\Delta_{\rm{expr}}$ & \we\ [cm$^{-1}$]  & $\Delta_{\rm{MRCI}}$  & $\Delta_{\rm{exp}}$  & \de \ [$\frac{kcal}{mol}$] &$\Delta_{\rm{MRCI}}$  &$\Delta_{\rm{exp}}$ \\\hline \hline

	oo-pCCD                   &  1.085 &($-0.019$)  &($-0.013$)  &2517 &($+176$) &($+158$) & 244.8 &($+26.9$)&($+19.7$)  \\ 
	
	pCCD-CID$'$               & 1.091 &($-0.013$)&($-0.007$) &2454  &($+113$) &($+95$) &304.3 &($+86.4$) &($+79.2$) \\ 

	pCCD-CID$'$+RDC           & 1.095 &($-0.009$) &($-0.003$)   &2407 & ($+66$)&($+48$) &263.3 &($+45.4$) &($+38.2$) \\ 
	pCCD-CID                 & 1.096 &($-0.008$) &($-0.002$)  &2399 & ($+58$)&($+40$) &266.1 & ($+48.2$) &($+41.0$)\\ 
	 pCCD-CID+RDC                  & 1.099  &($-0.005$)    &($+0.001$)          &2369 &($+28$)  &($+10$) &231.9 & ($+14.0$)&($+6.8$)   \\ 
	 fpLCCD                  &  1.101 &($-0.003$) &($+0.003$)           &2347 &($+6$) &($-12$)      &&&\\ 
	 fpCCD                  & 1.100   & ($-0.004$) &($+0.002$)          &2365 &($+24$) &($+6$)  &&&\\ 
	  CCD                  & 1.093 &($-0.011$) &($-0.005$)             &2447 & ($+106$)&($+88$)    &&&\\ 

	pCCD-CISD$'$                & 1.088 &($-0.016$) &($-0.013$) &2447 & ($+106$)&($+88$) &305.4 &($+87.5$)&($+80.3$)\\ 

	pCCD-CISD$'$+RDC             & 1.095 & ($-0.009$)&($-0.003$)  &2396 &($+55$) &($+37$) &263.3 &($+45.4$) &($+19.7$)\\ 

	pCCD-CISD                  & 1.097 &($-0.007$) &($-0.001)$ &2394 & ($+53$) &($+35$) &270.4 &($+52.5$)&($+45.3$)  \\ 
	 pCCD-CISD+RDC                  & 1.099 & ($-0.005$) &($+0.001$)           &2360 &($+19$) &($+1$)  &234.4 &($+16.5$) &($+9.3$)   \\ 
	 fpLCCSD                  &  1.102 &($-0.002$) &($+0.004$)            &2340 &($-1$)&($-19$) &&&  \\ 
	 fpCCSD                  & 1.100  &($-0.011$)   & ($+0.002$)           &2365 &($+24$) &($+6$) &&&\\ 
	  CCSD                  & 1.093   &($-0.011$) &($-0.005$)          & 2444 &($+103$) &($+85$)     &&& \\ \hline
	   MRCI~\cite{peterson1993-homo}                 &  1.104  && ($+0.006$)                &2341 & & ($-18$)           & 217.9 && ($-7.2$) \\ 
	   exp~\cite{herzberg_iv,shimanouchi}                   &  1.098  &($-0.006$)    &             &2359  &($+18$) &             & 225.1 &($+7.2$)&\\ 
\hline \hline
\end{tabular}
}
\end{table*}

\subsection{The Nitrogen Dimer}

Figure~\ref{fig:fig1} shows the PESs obtained from various pCCD-based methods and conventional single-reference CC models restricted to double and single and double excitations.
We performed pCCD-CID and pCCD-CISD calculations with and without a renormalized Davidson correction (RDC).
The corresponding results are indicated by including ``+RDC'' in the label.
Furthermore, the superscripts $^\prime$ indicates that the electron pair sector is included in CI calculations.

All investigated corrections generally predict a similar behavior around the equilibrium distances, while the shape of the PES features only small changes up to a bond distance of 1.6 \AA.
From this point onward, the PESs obtained from various pCCD-CI methods differ from those predicted by fp(L)CC as well as conventional CC approaches.
This observations is true for both variants, including only double (Figure~\ref{fig:fig1}-a) as well as single and double (Figure~\ref{fig:fig1}-b) excitations in the CI ansatz.
Specifically, for interatomic distances of $\ge2$ \AA, conventional methods as well as pCCD-based CC corrections diverge.
On the other hand, the potential energy curves predicted by pCCD-CI are smooth and do not suffer from divergencies along the reaction coordinate.
In the vicinity of dissociation (around 3 \AA{}), the simple RD correction, however, breaks down and yields undershooting PESs.
Nonetheless, a single-reference CI correction restricted to at most double excitations is able to provide a sound dissociation pathway for the \ce{N_2} molecule, while approaches based on CC corrections break down for stretched N--N bond lengths.

For a quantitative analysis, we determined the spectroscopic constants from the fitted PESs, which are summarized in Table~\ref{tbl:n2} and compared to MRCI and experimental reference data, respectively.
In general, all investigated CI and CC methods underestimate \re\ and overestimate \we\ and \de{}, respectively.
Both the CID and CISD correction provide similar spectroscopic constants.
Most importantly, including the electron pair sector in the CI ansatz considerably worsens vibrational frequencies and potential energy well depths, where the errors are approximately doubled and approach the CCD/CCSD level of accuracy.
Including a Davidson correction on top of pCCD-CID or pCCD-CISD significantly lowers errors in spectroscopic constants by more than a factor of 2.
In general, pCCD-CISD+RDC yields values for \re\ and \we\ that lie between fpLCCSD and fpCCD accuracy, while fpLCCSD predicts spectroscopic constants closest to MRCI reference data.
Conventional CC methods (CCD and CCSD) feature the largest deviations from MRCI and experimental reference data.
We should note that a CI correction excluding the electron pair sector, but including a Davidson correction, yields results closest to experiment (with differences up to 0.001 \AA{}, 1 cm$^{-1}$, or 9 kcal mol$^{-1}$ for \re{}, \we{}, and \de{}, respectively).
A significant advantage of pCCD-CI methods over CC-type corrections is the ability to dissociate the \ce{N2} molecule without encountering divergencies.
Similar to \re{} and \we{}, the most reliable results for \de{} are obtained from pCCD-CISD+RDC (that is, excluding the electron pair sector).
Although the fitted value of \de\ deviates by about 4\% from the experimental value, the absolute difference of 9 kcal mol$^{-1}$ is dissatisfying.
Compared to MRCI reference data, this error further increases to 16 kcal mol$^{-1}$.
In summary, pCCD-CI methods excluding the electron pair sector are more accurate than the corresponding CI flavours explicitly including electron pair excitations. 
Moreover, an RDC correction noticeably improves the performance of the proposed pCCD-CI models, significantly lowering the $\Delta_{\rm{MRCI}}$ and $\Delta_{\rm{exp}}$ errors.
In general, pCCD-CI outperforms conventional CC methods of similar cost, provides spectroscopic constants between the fpLCC and fpCC level of accuracy, and can efficiently model the static/nondynamical and dynamical correlation along the reaction coordinate. 

\begin{figure*}[t]
\hspace{-0.5cm}
	\includegraphics[width=2\columnwidth]{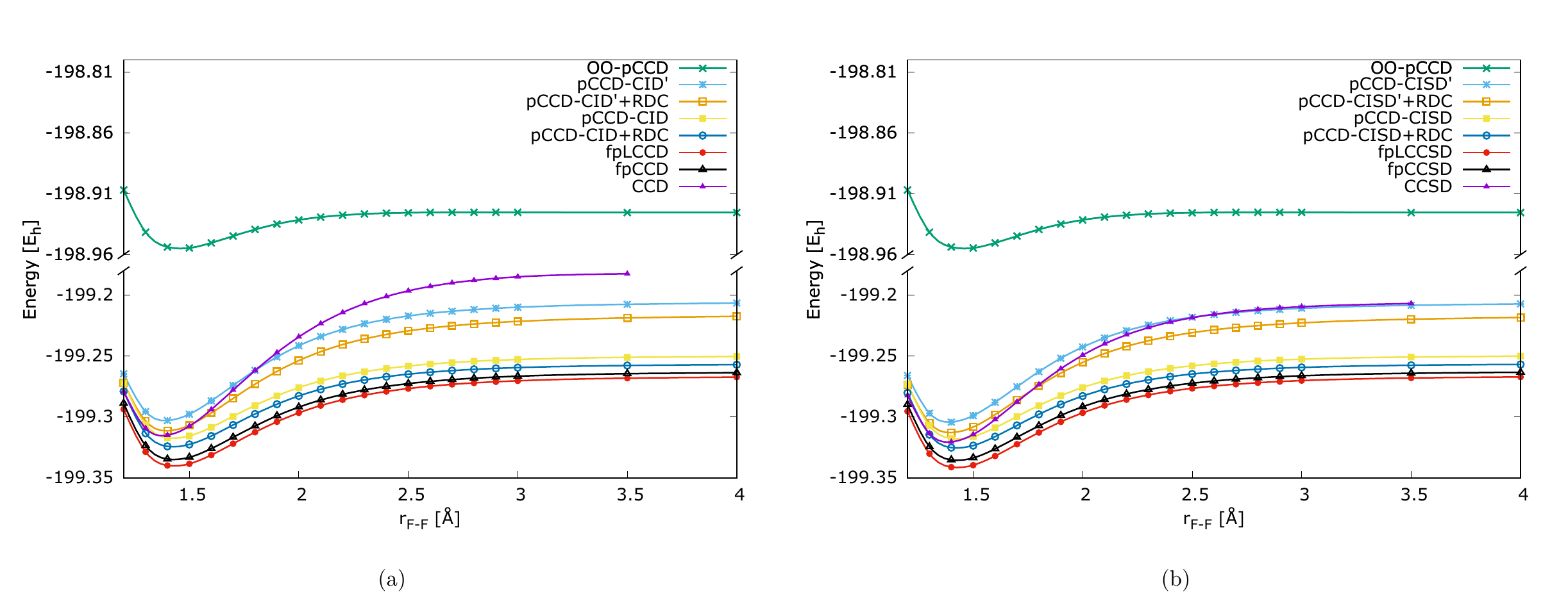}
	\tiny 
	\caption{Potential energy surfaces for the ground states of \ce{F_2} determined for various post-pCCD methods with (a) doubles excitation and (b) singles and doubles excitations. The superscript $^\prime$ indicates that the CI operator also includes the electron pair sector. RDC: renormalized Davidson correction.
	}
  	\label{fig:fig2}
\end{figure*}

\begin{table*}[tb]	
\centering
\caption{Spectroscopic parameters for the ground state of the \ce{F_2} molecule obtained from various post-pCCD methods and conventional CC models. \re\ is the equilibrium bond length, \we\ the harmonic vibrational frequency, and \de\ the dissociation energy, respectively.
The differences with respect to experimental result $\Delta_{\rm{exp}}$ or MRCI reference data $\Delta_{\rm{MRCI}}$ are given in parentheses. The total electronic energies are summarized in the SI.
RDC: renormalized Davidson correction.
} 
\label{tbl:f2}
{\footnotesize

\begin{tabular}{l|ccc|ccc|ccc}

	Method & \re\ [\AA]  & $\Delta_{\rm{MRCI}}$  & $\Delta_{\rm{expr}}$ & \we\ [cm$^{-1}$]  & $\Delta_{\rm{MRCI}}$  & $\Delta_{\rm{exp}}$  & \de \ [$\frac{kcal}{mol}$] &$\Delta_{\rm{MRCI}}$  &$\Delta_{\rm{exp}}$ \\\hline \hline

	oo-pCCD                   &  1.458 &($+0.038$)&($+0.046$) 	&733 &($-159$)&($-184$)	&18.6 &($-15.3$)&($-18.7$)\\ 
	
	pCCD-CID$'$     & 1.397 &($-0.023$) &($-0.015)$	&975 &($+83$)&($+58$)	&60.5 &($+26.6$)&($+22.8$) \\
	pCCD-CID$^\prime$+RDC    & 1.403 &($-0.017$) &($-0.009$)& 954  &($+62$)	&($+37$)	&59.0 &($+25.1$)&($+21.3$)  \\ 
	pCCD-CID    & 1.431 &($+0.011$)&($+0.019$)	&857 &($-35$)&($-60$)	&42.2 &($+8.3$) &($+4.5$)    \\ 
	 pCCD-CID+RDC   & 1.430 &($+0.010$)&($+0.018$) &857 &($-35$) &($-60$)	& 42.4 &($+8.5$)  &($+4.7$) \\
	fpLCCD         &  1.434 &($+0.014$) &($+0.022$)  &	866  & ($-26$)&($-51$)	& 44.3  &($+10.4$) &($+6.6$) \\ 
	fpCCD          &1.433 &($+0.013$)&($+0.021$)	& 863 &($-29$)&($-48$)	& 43.3 &($+9.4$)&($+5.6$)     \\ 
	  CCD            & 1.383 &($-0.037$) &($-0.029$) & 1053 &($+161$)&($+136$)	& 82.2 & ($+48.3$) &($+44.5$)   \\ 

	pCCD-CISD$^\prime$    &1.397 &($-0.023$)&($-0.015$)	&978 &($+86$)	&($+71$) 	&60.9 &($+27.0$)&($+23.2$)\\ 
	pCCD-CISD$'$+RDC    & 1.403  &($-0.017$)&($-0.009$)	&957 &($+65$)&($+40$)	&59.3 &($+25.4$) &($+21.6$)\\ 
	pCCD-CISD    &1.429 &($+0.009$)&($+0.017$)	&864 &($-28$)&($-53$)	&42.7 &($+8.8$ ) &($+5.0$)      \\
	 pCCD-CISD+RDC   &1.429 &($+0.009$)&($+0.017$)	&867 &($-25$)&($-50$)	&43.1 &($+9.2$)  &($+5.4$) \\ 
	 fpLCCSD        &  1.431 &($+0.011$)&($+0.019$)&	879 &($-13$)&($-38$)	&45.2 &($+11.3$)  &($+7.5$) \\ 
	 fpCCSD         & 1.431 &($+0.011$)&($+0.019$)	&873 &($-19$)&($-44$)  &	44.2 &($+10.3$) &($+6.5$)  \\ 
	  CCSD                     & 1.392 &($-0.028$)&($-0.020$)	 &1018 &($+126$)	&($+101$)	&70.1 &($+36.2$) &($+32.4$)  \\  \hline
	   MRCI~\cite{peterson1993-homo}           &  1.420  &&($+0.008$)     &892 &&($-25$)  &33.9 &&($-3.8$)  \\
	   exp~\cite{herzberg_iv,irikura2007}           & 1.412  &($-0.008$)&      &917 &($+25$)& &37.7&($+3.8$)& \\ 
\hline \hline
\end{tabular}
}
\end{table*}

\subsection{The Fluorine Dimer}

In contrast to the \ce{N_2} molecule, we do not observe any significant differences in the shape of the PESs predicted by all investigated pCCD corrections for the \ce{F2} dimer (see Figure~\ref{fig:fig2}).
The similarities in the shape of the potential energy curves translate into similar values of spectroscopic constants, which are summarized in Table~\ref{tbl:f2}.
As expected, major differences ($\Delta_{\rm{MRCI}}$) are obtained if the electron pair sector is included in the CI ansatz.
We should note that, in contrast to the \ce{N2} dimer, $\Delta_{\rm{exp}}$ for \re\ and \we\ significantly decreases in CI$^\prime$ corrections, while errors in \de\ considerably increase.
Most importantly, the changes in spectroscopic constants are only minor if single excitations are included in the CI or CC correction.
Moreover, the accuracy in spectroscopic constants is similar for pCCD-CI methods (without electron pair excitations) and fp(L)CC models.
While pCCD-CI provides smaller errors in \re\ and \de{}, fp(L)CC yields values of \we\ that are closer to MRCI reference data.
Specifically, both types of corrections overestimate \re\ by around 0.02 (0.01) \AA\ compared to experimental (MRCI) reference data, while the differences in \de\ amount to approximately 10 kcal mol$^{-1}$.
Furthermore, the proposed pCCD-CI methods increase the error in harmonic vibrational frequencies \we\ by about 10 cm$^{-1}$ with respect to fp(L)CC.
Including a RD correction on top of the pCCD-CI models has only a minor effect on spectroscopic constants.
While \re\ remains unchanged, \we\ and \de\ only marginally worsen by 3 cm$^{-1}$ and 0.4 kcal mol$^{-1}$.
Finally, all pCCD corrections outperform the conventional CCD or CCSD models, which provide large errors in \re\ (about 0.03 \AA{}), \we\ (126 cm$^{-1}$), and \de\ (about 36 kcal mol$^{-1}$).

To sum up, the quality and performance of the investigated CI corrections on top of a pCCD reference function is comparable to fp(L)CC methods featuring a similar excitation order.
Nonetheless, for problems which feature a substantial amount of static electron correlation effects (like the dissociation process of \ce{N2}), the proposed pCCD-CI models seem more robust and less dependent on the quality of the pCCD reference.
Thus, pCCD-CI might be less prone to unexpected failures encountered in CC corrections like divergencies.

\section{Assessing the accuracy of CI corrections in actinide-containing compounds}\label{sec:exstates}

In the following, we extend our test set of light diatomic molecules with small di- and triatomic actinide compounds.
Specifically, we investigate the simplest actinide compounds, ThO~\cite{Marian_tho_1988,theo_casscf_paulovic,PhysRevA.78.010502,edm_fleig_nayak2014,Skripnikov-ThO-2015,Skripnikov-ThO-2016,Denis-ThO-2016,ptThS} and ThS~\cite{ThS_liang_andrews_2002,ptThS}, which are important systems in studies on the electron electric dipole moment (eEDM)~\cite{heaven:barker:antonov,wang:le:steimle:heaven,ACME_science,Sulfur_Heaven_2014,ptThS}.
Furthermore, we scrutinize the performance of the CI corrections for the the \ce{UO2^{2+}} isoelectronic series~\cite{Wadt1981,dyall_1999,uranyl_de_jong_99,uranyl_pitzer_99,kaltsoyannis,matsika_2001,real07,real09,pawel1,fen_wei,pawel2,pawel3,pawel_saldien,tecmer2014,gomes2015applied,gomes_crystal,tecmer-song2016}, containing \ce{UO2^{2+}}, \ce{UN2}, \ce{PaO2^{+}} and \ce{ThO_{2}}~\cite{denning2007,baker2012,xuy,pierloot05,real09,nun,kovacs2011,pawel2,kovacs-chem-rev-2015,tecmer-song2016}.
Finally, we augment our numerical analysis to target also the lowest-lying excited states in the \ce{UO2^{2+}} cation.

\begin{table*}[tb]	
\centering
\caption{Spectroscopic parameters of the ground states of the investigated actinide-containing compounds obtained from various post-pCCD methods and conventional CC methods. \re\ is the equilibrium bond length, \we\ the harmonic vibrational frequency,  respectively.
 The differences with respect to CCSD(T) are given in parentheses. The total electronic energies, as also PES curves are summarized in the ESI\dag.
 RDC: renormalized Davidson correction.}
\label{tbl:actinide}
{\footnotesize
\begin{tabular}{l|cc|cc|cc}
	Method & \re\ [\AA]  &~~ \we\ [cm$^{-1}$] & \re\ [\AA]  &~~ \we\ [cm$^{-1}$]  & \re\ [\AA]  &~~ \we\ [cm$^{-1}$]   \\\hline \hline 
    & \multicolumn{2}{c|}{ThO} & \multicolumn{2}{c}{ThS}& \multicolumn{2}{c|}{$\ce{UO_2}^{2+}$}\\ \cline{1-7}

	oo-pCCD                & 1.854 ($-0.012$)   &~~ 894 ($+56$) & 2.387 ($+0.011$) &~~ 484 ($+12$)  & 1.670 ($-0.038$) &~~ 1139 ($+72$) \\
	pCCD-CID$^\prime$      & 1.850 ($-0.016$)   &~~ 898 ($+60$) & 2.365 ($-0.011$) &~~ 494 ($+22$)  & 1.681 ($-0.027$) &~~ 1164 ($+97$) \\
	pCCD-CID$^\prime$+RDC  & 1.853 ($-0.013$)   &~~ 890 ($+52$) & 2.366 ($-0.010$) &~~ 492 ($+20$)  & 1.687 ($-0.021$) &~~ 1154 ($+87$) \\
	pCCD-CID               & 1.854 ($-0.012$)   &~~ 888 ($+50$) & 2.366 ($-0.010$) &~~ 494 ($+22$)  & 1.687 ($-0.021$) &~~ 1129 ($+62$) \\ 
	 pCCD-CID+RDC          & 1.855 ($-0.011$)   &~~ 886 ($+48$) & 2.365 ($-0.011$) &~~ 493 ($+21$)  & 1.690 ($-0.018$) &~~ 1114 ($+47$) \\ 
	 fpLCCD                & 1.858 ($-0.008$)   &~~ 876 ($+38$) & 2.367 ($-0.009$) &~~ 488 ($+16$)  & 1.701 ($-0.007$) &~1067 ($\pm 0$)\\ 
	 fpCCD                 & 1.856 ($-0.010$)   &~~ 869 ($+31$) & 2.365 ($-0.011$) &~~ 491 ($+19$)  & 1.698 ($-0.010$) &~~ 1087 ($+20$)  \\ 
	  CCD                  & 1.851 ($-0.015$)   &~~ 895 ($+57$) & 2.362 ($-0.014$) &~~ 492 ($+20$)  & 1.683 ($-0.025$) &~~~~ 1174 ($+107$) \\
	pCCD-CISD$^\prime$     & 1.852 ($-0.014$)   &~~ 894 ($+56$) & 2.368 ($-0.008$) &~~ 494 ($+22$)  & 1.683 ($-0.025$) &~~ 1158 ($+91$) \\
	pCCD-CISD$^\prime$+RDC & 1.856 ($-0.010$)   &~~ 889 ($+51$) & 2.370 ($-0.006$) &~~ 487 ($+15$)  & 1.689 ($-0.019$) &~~ 1129 ($+62$) \\
	pCCD-CISD              & 1.855 ($-0.011$)   &~~ 889 ($+51$) & 2.367 ($-0.009$) &~~ 492 ($+20$)  & 1.689 ($-0.019$) &~~ 1125 ($+58$) \\ 
	pCCD-CISD+RDC          & 1.857 ($-0.009$)   &~~ 871 ($+33$) & 2.369 ($-0.007$) &~~ 487 ($+15$)  & 1.694 ($-0.014$) &~~ 1102 ($+35$) \\ 
	 fpLCCSD               & 1.863 ($-0.003$)   &~840 ($+2$)    & 2.384 ($+0.008$) &~~ 457 ($-15$)  & 1.717 ($+0.009$) &~~~ 976 ($-91$) \\ 
	 fpCCSD                & 1.857 ($-0.009$)   &~~ 874 ($+36$) & 2.367 ($-0.009$) &~~ 489 ($+17$)  & 1.698 ($-0.010$) &~~ 1115 ($+48$) \\ 
	  CCSD                 & 1.855 ($-0.011$)   &~~ 885 ($+47$) & 2.366 ($-0.010$) &~~ 487 ($+15$)  & 1.690 ($-0.018$) &~~ 1142 ($+75$) \\ 
	   CCSD(T)             & 1.866              & 838           & 2.376            &~~ 472          & 1.708            &~~ 1067   \\ \hline
  & \multicolumn{2}{c|}{$\ce{ThO_2}$} & \multicolumn{2}{c}{$\ce{PaO_{2}^{+}}$}
  & \multicolumn{2}{c}{$\ce{UN_2}$} \\ \cline{1-7}

	oo-pCCD                & 1.917 ($-0.012$)  & 805 ($-13$)  & 1.763 ($-0.021$) & 972 ($\pm 0$)  & 1.711 ($-0.030$) &~~1176 ($+74$)  \\
	pCCD-CID$^\prime$      & 1.913 ($-0.016$)  & 858 ($+40$)  & 1.763 ($-0.021$) & 1024 ($+52$)   & 1.717 ($-0.024$) &~~1171 ($+69$)  \\
	pCCD-CID$^\prime$+RDC  & 1.917 ($-0.012$)  & 841 ($+23$)  & 1.767 ($-0.017$) & 1011 ($+39$)   & 1.722 ($-0.019$) &~~1164 ($+62$)  \\
	pCCD-CID               & 1.917 ($-0.012$)  & 852 ($+34$)  & 1.768 ($-0.016$) & 1000 ($+28$)   & 1.720 ($-0.021$) &~~1124 ($+22$)  \\ 
	 pCCD-CID+RDC          & 1.918 ($-0.011$)  & 850 ($+32$)  & 1.771 ($-0.013$) &~ 987 ($+15$)   & 1.725 ($-0.016$) &~~1122 ($+20$)  \\ 
	 fpLCCD                & 1.925 ($-0.004$)  & 829 ($+11$)  & 1.778 ($-0.006$) &~ 995 ($+23$)   & 1.736 ($-0.005$) &~~1077 ($-25$)  \\ 
	 fpCCD                 & 1.924 ($-0.005$)  & 830 ($+12$)  & 1.776 ($-0.008$) &~ 985 ($+13$)   & 1.731 ($-0.011$) & 1102 ($\pm 0$) \\ 
	  CCD                  & 1.914 ($-0.015$)  & 850 ($+32$)  & 1.765 ($-0.019$) & 1027 ($+55$)   & 1.722 ($-0.019$) &~~1184 ($+82$)  \\
	pCCD-CISD$^\prime$     & 1.914 ($-0.015$)  & 851 ($+33$)  & 1.765 ($-0.019$) & 1014 ($+42$)   & 1.719 ($-0.022$) &~~1162 ($+40$)  \\
pCCD-CISD$^\prime$+RDC     & 1.918 ($-0.011$)  & 842 ($+24$)  & 1.770 ($-0.014$) & 1004 ($+32$)   & 1.725 ($-0.016$) &~~1152 ($+50$)  \\
	pCCD-CISD              & 1.918 ($-0.011$)  & 837 ($+19$)  & 1.770 ($-0.014$) &~ 999 ($+27$)   & 1.722 ($-0.019$) &~~1118 ($+16$)  \\ 
	pCCD-CISD+RDC          & 1.920 ($-0.009$)  & 836 ($+18$)  & 1.774 ($-0.010$) &~ 982 ($+10$)   & 1.727 ($-0.014$) & 1106 ($+4$)    \\ 
	 fpLCCSD               & 1.934 ($+0.005$)  & 773 ($-45$)  & 1.789 ($+0.005$) &~ 895 ($-77$)   & 1.748 ($+0.007$) &~~~~~ 960 ($-142$) \\
	 fpCCSD                & 1.923 ($-0.006$)  & 829 ($+11$)  & 1.776 ($-0.008$) &~ 989 ($+17$)   & 1.733 ($-0.008$) &~~1113 ($+11$)  \\
	  CCSD                 & 1.918 ($-0.011$)  & 804 ($-14$)  & 1.770 ($-0.014$) &~ 965 ($-75$)   & 1.727 ($-0.014$) &~~1158 ($+56$)  \\
	   CCSD(T)             & 1.929             & 818          & 1.784            & 972            & 1.741            & 1102\\ \hline

\hline \hline

\end{tabular}

}

\end{table*}

\subsection{Targeting the ground states of actinide compounds}

The spectroscopic constants for the electronic ground states of the investigated actinide compounds are summarized in Table~\ref{tbl:actinide} and are obtained from the corresponding PESs, which are collected in the SI.
For \ce{ThO}, the inclusion of single excitations on top of doubles in the correction significantly improves the equilibrium distance \re{}. 
This is a common feature for all investigated pCCD corrections, where fpLCCSD yields the most accurate values, which differ only about 0.003 \AA\ from the CCSD(T) reference results. 
fpCCSD and pCCD-CISD+RDC (that is, including a RD correction) perform similarly, both underestimating \re\ by about 0.009 \AA{}. 
Similar observation can be made for the harmonic vibrational frequencies.
Specifically, the LCCSD correction approaches spectroscopic accuracy (deviating by 2 cm$^{-1}$ from the reference), while fpCCSD and pCCD-CISD+RDC feature larger differences of about 36 and 33 cm$^{-1}$, respectively.

In case of the ThS molecule, all investigated pCCD corrections yield spectroscopic constants of similar accuracy.
While all CI corrections and the conventional CCSD model underestimate \re{}, the LCCSD correction overestimates equilibrium bond lengths. 
In absolute value, all corrections provide similar error measures amounting from 0.007 to 0.011 \AA\ with respect to CCSD(T) reference values. 
The fitted values for \we\ are very consistent, where all corrections result in an error of approximately 20 cm$^{-1}$ with respect to the CCSD(T) harmonic vibrational frequency.
As observed above, adding a RD correction generally improves the performance of all pCCD-CI-type approaches. 
However, we should note that the pCCD approximation already delivers a well description of spectroscopic constants for both the ThO and ThS molecules.
All applied corrections do not significantly improve the pCCD results. 

Quantum chemistry studies on the \ce{UO2^{2+}} isoelectronic series are still an active field of research. 
These small triatomic molecules have a linear geometry beyond the \ce{ThO_2} species, which features a bent geometry and a peculiar electronic structure. 
In contrast to its linear analogues, the Thorium 5f orbitals of \ce{ThO_2} participate neither in bonding nor in electronic excitations. 
This is caused by differences in the relative energetic ordering of the 5f and 6d orbitals in the thorium and uranium atoms, respectively.
By scrutinizing the spectroscopic constants summarized in Table \ref{tbl:actinide}, we are able to assess whether the investigated post-pCCD methods can reliably model electron correlation effects in these compounds along the PESs. 
Starting from the uranyl cation, pCCD considerably underestimates the equilibrium distance \re{}.
All CI corrections significantly improve this value decreasing the errors by a factor of 2.
Specifically, pCCD-CISD+RDC underestimate \re\, by $-0.014$ \AA{}. 
A better performance can be observed for fp(L)CC methods, which differ by only $0.007-0.009$ \AA{} in absolute value. 
Predicting accurate \we\ is more challenging.
In order to reduce the pCCD errors by a factor of 2, a CISD+RDC correction is required, while the remaining CI flavours feature large differences with respect to CCSD(T) reference values (more than 60 cm$^{-1}$).
The performance of the fp(L)CC methods is more erratic.
While fpLCCD predicts a \we\ that perfectly agrees with the CCSD(T) reference, inclusion of single excitations leads to large errors (up to 90 cm$^{-1}$).
Moving to a fpCCSD formalism improves spectroscopic constants, where the errors in \we\ amount to 48 cm$^{-1}$.
We should highlight that the conventional CCSD method provides larger errors in spectroscopic constants than the alternatives fp(L)CCSD or pCCD-CISD+RDC.

Analogous to the examples above, the CI corrections significantly improves pCCD results for the \ce{UN2} molecule, where pCCD-CISD+RDC provides the smallest errors with respect to CCSD(T) reference data. 
Furthermore, we observe the same tendency for all investigated corrections to underestimate the equilibrium bond length and to overestimate the harmonic vibrational frequency, except for fpLCCSD, which shows the opposite behaviour. 
Specifically, the (L)CCSD corrections provide more accurate equilibrium bond lengths, while the CISD flavours reduce the errors in \we\ to 4 cm$^{-1}$.
Most importantly, fpCCSD and pCCD-CISD+RDC outperform the conventional CCSD approach and predict spectroscopic constants that deviate the least from CCSD(T) reference values.

For the protactinium dioxygen cation, we observe similar trends, where the CI/CC corrections are able to cure some of the deficiencies originated from the pCCD model.
Specifically, the errors in \re\ are reduced by a factor of 2 in CI-type corrections (pCCD-CISD+RDC features an error of 0.01 \AA{}) and further lowered to 0.005 \AA{} in the LCC corrections.
However, all investigated pCCD corrections deteriorate harmonic vibrational frequencies, where pCCD already provides the proper shape of the PESs around the equilibrium distance.
While pCCD-CISD+RDC yields the smallest errors of 10 cm$^{-1}$, fpLCCSD increases the discrepancy to 77 cm$^{-1}$.
Nonetheless, both CID/CISD and (L)CCD/(L)CCSD corrections again outperform the conventional CCD/CCSD approaches, which feature one of the largest errors in spectroscopic constants. 

A particular difficult case is the \ce{ThO2} molecule, where the CI corrections do not significantly change or improve the spectroscopic constants derived from the pCCD PES.
The best performance is obtained from the fp(L)CC methods, where the errors in \re\ are reduced to 0.005 \AA{}, while the absolute errors in \we\ remain unchanged (except for fpLCCSD, which provides the largest errors).
The most accurate CI flavor is again pCCD-CISD+RDC, which underestimates \re\ by around 0.009 \AA, while \we\ deviates by 18 cm$^{-1}$.
The performance of the CI corrections lies between fp(L)CC and conventional CC theory.

\subsection{Dissecting the accuracy of pCCD corrections based on an error analysis.}

To scrutinize the accuracy of the investigated corrections in predicting accurate values for \re\ and \we{}, we perform a statistical analysis.
Specifically, we evaluate the mean error (ME) of the fitted spectroscopic constants with respect to CCSD(T) reference data,  
\begin{equation}\label{eq:me}
    \mathrm{ME}=\sum_{i}^{N}\Delta x_{i}/N,
\end{equation}
where $\Delta x_{i}$ is the difference in \re\ or \we\ between a selected method and the reference $\Delta x_i = x_i^{\rm method} - x_i^{\rm CCSD(T)}$.
The sum runs over the all investigated actinide-containing compounds ($N=6$).
Furthermore, we also determine the root-mean-square deviation (RMSD) in \re\ and \we{}, according to
\begin{equation}\label{eq:rmsd}
    \mathrm{RMSD}=\sqrt{\sum_{i}^{N}\Delta x_{i}^2/N}.
\end{equation}

\begin{figure}[t]
\centering
\includegraphics[width=1.0\columnwidth]{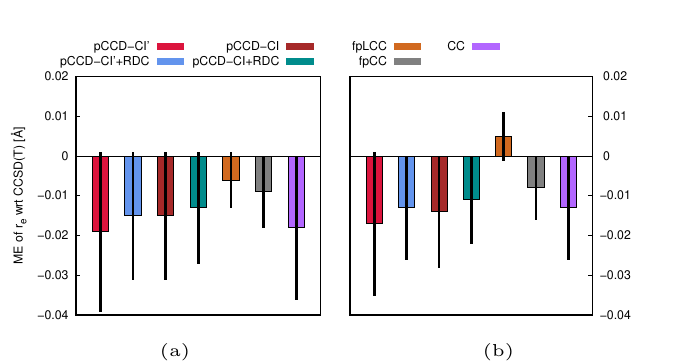}
\centering 
\scriptsize
\caption{Mean errors in \re\ obtained for various post-pCCD and conventional CC methods restricted to (a) double excitations and (b) single and double excitations with respect to CCSD(T) reference data calculated from eq.~\ref{eq:me} for all investigate actinide-containing compounds. The standard deviation (RMSE) is indicates as black lines.
The corresponding numerical values are summarized in Table S19 in the SI.
RDC: renormalized Davidson correction.
}
\label{fig:error1}
\end{figure}

The graphical representation of the ME and RMSD of \re\ is shown in Figure~\ref{fig:error1} for CI and CC methods including (a) only double and (b) single and double excitations.
On average, all investigated methods underestimates the equilibrium bond distance.
The smallest deviations from CCSD(T) data are obtained from fpLCCD/fpLCCSD, followed by fpCCD/fpCCSD and then by pCCD-CID+RDC/pCCD-CISD+RDC.
The largest discrepancies are deduced from the pCCD-CID$^\prime$/pCCD-CISD$^\prime$ PESs.
These large errors might be related to the double-counting problem of electron correlation effects associated with the electron pairs, which are included in both the pCCD reference function and the \textit{a posteriori} CI correction.
Excluding pair excitations in the CI operator significantly reduces errors.
Furthermore, including single excitations in the pCCD correction only slightly affects equilibrium bond lengths, where the ME and RMSE are reduced by 0.002 \AA{} in CI corrections and by 0.001 \AA{} in the (L)CC flavors.
However, the performance of fpLCC methods is more erratic; while fpCCD underestimates \re{}, fpCCSD overestimates equilibrium distances.
Finally, we should note that all pCCD corrections (excluding the pair sector and including a RD correction) provide smaller ME and RMSE compared to the conventional CC methods featuring similar excitation operators.

\begin{figure}[t]
\centering
\includegraphics[width=1.0\columnwidth]{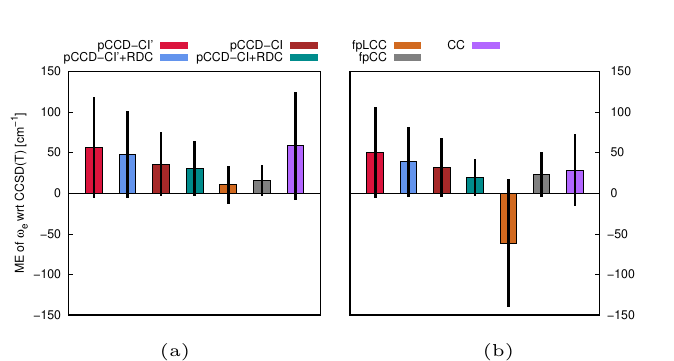}
\centering 
\scriptsize
\caption{Mean errors in \we\ obtained for various post-pCCD and conventional CC methods restricted to (a) double excitations and (b) single and double excitations with respect to CCSD(T) reference data calculated from eq.~\ref{eq:me} for all investigate actinide-containing compounds. The standard deviation (RMSE) is indicates as black lines.
The corresponding numerical values are summarized in Table S20 in the SI.
RDC: renormalized Davidson correction.
}
\label{fig:error2}
\end{figure}

Figure~\ref{fig:error2} highlights the ME and RMSD for \we{} for all CI and CC methods restricted to (a) double excitations and (b) single and double excitations.
A similar behaviour in ME and RMSD can be be observed as found for \re{}.
Considering only double excitations, fpLCCD yields the smallest ME of 10 cm$^{-1}$, followed by fpCCD with a ME up to 20 cm$^{-1}$, while pCCD-CID+RDC features an increased ME of about 30 cm$^{-1}$.
All investigated approaches tend to overestimate the harmonic vibrational frequency.
Note that the largest errors are obtained for the conventional CCD model.
Including both single and double excitations significantly changes the distribution of ME and RMSE.
The smallest ME is obtained for pCCD-CISD+RDC (19 cm$^{-1}$), followed by fpCCSD with a ME of about 23 cm$^{-1}$.
Furthermore, the inclusion of single excitations in the correction significantly increases the accuracy of all CI-type corrections, while the ME of fpCC methods grows by about 70\%.
Similarly, the performance of fpLCC considerably deteriorates raising ME and RMSE by a factor of 6 and 4, respectively.
To summarize, fpLCCSD provides the smallest errors in \re{}, while pCCD-CISD+RDC properly describes the shape of the PESs yielding the most accurate harmonic vibrational frequencies.

\begin{table}[b]
\scriptsize
\caption{Spectroscopic constants of some low-lying singlet excited states of \ce{UO2^{2+}} obtained from various post-pCCD methods and conventional EOM-CC theory. \re\ is the equilibrium bond length, \we\ the harmonic vibrational frequency, \Ta\ the adiabatic excitation energy, and \Tv\ the vertical excitation energy, respectively.
\Tv\ has been calculated with respect to the equilibrium bond distance \re\ of the corresponding ground state obtained for each method. The differences with respect to CR-EOM-CCSD(T) are given in parentheses.
The superscripts $^\prime$ indicates that the pCCD-CISD method includes the electron pair sector in the excitation operator.
EOM-fpLCCSD was originally introduces as EOM-pCCD-LCCSD~\cite{eom-pccd-lccsd}.
}
\centering
\label{tbl:u1u2}
{
\begin{tabular}{l|cccc}
    & \multicolumn{4}{c}{\ce{UO_2^{2+}}}\\ \cline{2-5}
	Method & \re\ [\AA]  & \we\ [cm$^{-1}$]&\Ta\ [eV] & \Tv [eV] \\\hline 
	& \multicolumn{4}{c}{$1^1\Phi_g(\sigma_u\rightarrow\phi_u)$}   \\ \hline 
pCCD-CISD$^\prime$ & 1.773($+$0.018)  & 888($-$49)  &14.50($+$10.74) &14.92($+$10.93) \\ 
	pCCD-CISD      & 1.773($+$0.018)  & 888($-$49)  &14.50($+$10.74) &14.92($+$10.93) \\ 
	EOM-fpLCCSD    & 1.786($+$0.031)  & 837($-$100) & 3.66($-$0.10)  & 3.87($-$0.12)\\
	EOM-CCSD       & 1.762($+$0.007)  & 924($-$13)  & 3.57($-$0.19)  & 3.86($-$0.13) \\ 
	CR-EOM-CCSD    & 1.755            & 937         & 3.76           & 3.99 \\  \hline
	& \multicolumn{4}{c}{$1^1\Delta_g(\sigma_u\rightarrow\delta_u)$}   \\ \hline
	pCCD-CISD$'$   & 1.791($+$0.039)  & 764($-169$) &14.72($+$10.56) &15.17($+$10.80) \\ 
	pCCD-CISD      & 1.791($+$0.039)  & 764($-169$) &14.72($+$10.56) &15.17($+$10.80)  \\
	EOM-fpLCCSD    & 1.781($+$0.029)  & 840($-$93)  &4.07($-$0.09)   & 4.25($-$0.12)\\
	EOM-CCSD       & 1.760($+$0.008)  & 910($-$23)  & 3.96($-$0.20)  & 4.22($-$0.15) \\ 
	CR-EOM-CCSD    & 1.752            & 933         & 4.16           & 4.37\\ \hline 
	
		& \multicolumn{4}{c}{$2^1\Phi_g(\pi_u\rightarrow\delta_u)$}   \\ \hline
pCCD-CISD$^\prime$ & 1.803($+$0.013)  & 928($+22$)  &14.75($+$9.85)  &15.34($+$9.89) \\ 
	pCCD-CISD      & 1.801($+$0.012)  & 975($+69$)  &14.74($+$9.84)  &15.34($+$9.89)  \\ 
	EOM-fpLCCSD    & 1.804($+$0.014)  & 941($+$35)  & 4.87($-$0.03)  & 5.29($-$0.16)\\
	EOM-CCSD       & 1.797($+$0.007)  & 902($-$4)   & 4.46($-$0.44)  & 5.08($-$0.37)\\
	CR-EOM-CCSD(T) & 1.790            & 906         & 4.90           & 5.45 \\ \hline
	& \multicolumn{4}{c}{$1^1\Pi_g(\pi_u\rightarrow\delta_u)$}   \\ \hline
	pCCD-CISD$'$   & 1.784($-$0.006)  & 1023($+118$) &14.83($+$9.84)  &15.53($+$10.00) \\ 
	pCCD-CISD      & 1.785($-$0.005)  & 1042($+137$) &14.82($+$9.83)  &15.52($+$9.99)  \\ 
	EOM-fpLCCSD    & 1.804($+$0.014)  &  965($+$60)  & 4.98($-$0.01)  & 5.38($-$0.15)\\
	EOM-CCSD       & 1.796($-$0.006)  &  904($-$1)   & 4.55($-$0.44)  & 5.16($-$0.37)\\
	CR-EOM-CCSD(T) & 1.790            &  905         & 4.99           & 5.53  \\
\hline \hline
\end{tabular}
}
\end{table}
\subsection{Performance of pCCD-CI methods for excited states.}

Finally, we will turn our discussion toward electronically excited states and scrutinize the performance of the investigated pCCD-CI methods to model the lowest-lying electronic states in the \ce{UO_2^{2+}} molecule.
Specifically, we target the following electronically excited states: $\sigma_u \rightarrow \delta_u$, $\sigma_u   \rightarrow \phi_u$ and two $\pi_u\rightarrow\delta_u$ transitions.
Or analysis is based on our previous study~\cite{nowak-pccp-2019}, where we focused on dissecting the accuracy of the LCC corrections in targeting electronically excited states in selected actinide compounds.
As a reference method, we chose the completely reonormalized equation of motion coupled cluster singles doubles and perturbative triples (CR-EOM-CCSD(T)) method.~\cite{cr-eomccsd}

Table~\ref{tbl:u1u2} contains the fitted spectroscopic constants for the four lowest-lying targeted excited states.
pCCD-CISD with and without electron pair excitations yield equivalent results (for \re\ and excitation energies) due to the purely singly-excited character of the targeted states.
However, the excitation energies predicted by the CI corrections strongly deviate from all EOM-CC results.
While the ground state energies are corrected towards lower energies (by accounting for the missing dynamical correlation energy), all electronically excited states miss this shift in energy.
Although pCCD-CISD provides reasonable excitation energies with respect to the pCCD reference energy, it misses a large fraction of the dynamical correlation energy of the excited states.
The adiabatic and vertical excitation energies collected in Table~\ref{tbl:u1u2} demonstrate that this shift in energy between pCCD-CISD and the CCSD/CR-EOM-CCSD(T) reference amounts to over 10 eV.
This disproportion in excitation energies is significant, especially compared to EOM-fpLCCSD, which differs only around from 0.1 to 0.5 eV from reference values.
Although these quantitative deficiencies of pCCD-CISD methods in capturing electron correlation effects in excited states forbid us to precisely target excitation energies, the CI corrections predict equilibrium bond lengths with reasonable accuracy providing smaller differences from CR-EOM-CCSD(T) reference data than EOM-fpLCCSD.
The prediction of harmonic vibrational frequencies is, however, more erratic, where errors lie between 12 and 140 cm$^{-1}$.
Specifically, for the lowest excited state $1^1\Phi_g$, pCCD-CISD predicts spectroscopic constants that are closer to reference values than the corresponding EOM-fpLCCSD results.
Thus, pCCD-CISD generally fails in describing electronically excited states and provides spectroscopic constants of irregular accuracy.
Similar disadvantages of CI corrections in predicting excitation energies were noticed in previous numerical studies~\cite{Kallay1}.

\section{\label{sec:conclusions}Conclusions}

In this work, we developed various CI corrections with a pCCD references function that are restricted to at most double (hole-particle) excitations.
Furthermore, we tested CI models, where the linear CI operator excludes and includes electron pair excitations, where the former case might be important to properly describe doubly-excited states.
The accuracy of the proposed pCCD-CI approaches was assessed for the ground states of light diatomic molecules, namely \ce{F_2} and \ce{N_2}, as well as for selected di- and tri-atomic f0 actinide species, such as the ThO, ThS, \ce{UO^{2+}_2}, \ce{NUN}, \ce{PaO^{+}_2}, and \ce{ThO_2} molecules.
We also scrutinized the performance of the CI corrections on the lowest-lying electronically excited states in the \ce{UO^{2+}_2} cation. 
The CI corrections were compared to fp(L)CC methods as well as to conventional CC variants.
Finally, we introduced a simple renormalized Davidson correction to minimize size-consistency errors in the ground state electronic structures.

Our study demonstrates that the proposed pCCD-CI methods provide a reliable and promising alternative to conventional CC methods as well as to the other unconventional approaches such as the fp(L)CC model.
A major advantage of the CI-type corrections is their ability to dissociate the \ce{N2} molecule, where the contributions to electron correlation effects change along the reaction coordinate.
The accuracy of the pCCD-CI models can be improved by including an (approximate) renormalized Davidson correction, where the errors in spectroscopic constants are reduced by a factor of 2.
In general, pCCD-CISD+RDC allows us to obtain smooth PESs and predicts spectroscopic constants that agree well with experimental or MRCI reference data.
Specifically, pCCD-CISD+RDC outperforms conventional CCSD, while the spectroscopic constants lie between fpLCCSD and fpCCSD accuracy. 
The good performance of pCCD-CISD+RDC is also observed for heavy-element-containing systems, where it features the smallest ME and RMSE in \we{}, while the corresponding error measures for \re\ are slightly smaller than the CCSD values.
The inclusion of electron-pair excitations in the CI operator, however, deteriorates the performance of the various CI corrections, which might originate from a double-counting problem of correlation effects associated with electron pair states as they are included in both the pCCD reference function and the \textit{a posteriori} CI correction.

Finally, the pCCD-CISD model fails for electronically excited states, where large errors in excitation energies are observed.
Although pCCD-CISD provides more accurate values for \re\ in electronic excited states than the EOM-fpLCCSD model, the prediction of \we\ and (adiabatic and vertical) excitation energies is rather erratic.
To sum up, a CI correction (including a Davidson correction) on top of pCCD is a promising alternative to conventional (like CCSD) or unconventional (like fp(L)CC) CC methods to model electronic ground states.
However, in order to reach chemical or spectroscopic accuracy, further improvements are required, like approximate protocols to account for higher-order excitations.
Furthermore, in this work, the CI operators are restricted to singlet excitations, while the projection manifold has been defined to allow for an algebraic spin-summation.
By eliminating this constraint, the CI corrections would allow us to target different spin sectors.
If we restrict the excitation operators to be of SD-type, both singlet and triplet states will be accessible in the CI corrections.
The corresponding diagonalization problem has to be performed in the Slater determinant basis containing all singly- and doubly-excited determinants with respect to $\ket{\Phi_0}$ instead of the spin-summed states defined in eqs.~\eqref{eq:projection-s} and \eqref{eq:projection-d}.
This formulation, however, increases the number of degrees of freedom of the diagonalization problem.

\section{Data Availability Statement}
The data that supports the findings of this study are available within the article and its supplementary material.

\begin{acknowledgments}
A.N.~and K.B.~acknowledge financial support from a SONATA BIS grant of the National Science Centre, Poland (no.~2015/18/E/ST4/00584).
A.N.~also received financial support from a PRELUDIUM 17 grant of the National Science Centre, Poland (no.~2019/33/N/ST4/01880).
\end{acknowledgments}

\normalem
\bibliography{rsc}

\end{document}